\documentclass[twocolumn,showpacs,preprintnumbers,amsmath,amssymb]{revtex4}

\usepackage{graphicx}
\usepackage{dcolumn}
\usepackage{bm}
\usepackage{epsfig}

\begin{document}

\title{Dielectric and thermal relaxation in the energy landscape}

\author{U. Buchenau}
\email{buchenau-juelich@t-online.de}
\author{R. Zorn}
\author{M. Ohl}
\author{A. Wischnewski}
\affiliation{Institut f\"ur Festk\"orperforschung, Forschungszentrum J\"ulich\\
Postfach 1913, D--52425 J\"ulich, Federal Republic of Germany}
\date{April 30, 2006; revised July 3, 2006}

\pacs{64.70.Pf, 77.22.Gm}

\begin{abstract}
We derive an energy landscape interpretation of dielectric relaxation times in undercooled liquids, comparing it to the traditional Debye and Gemant-DiMarzio-Bishop pictures. The interaction between different local structural rearrangements in the energy landscape explains qualitatively the recently observed splitting of the flow process into an initial and a final stage. The initial mechanical relaxation stage is attributed to hopping processes, the final thermal or structural relaxation stage to the decay of the local double-well potentials. The energy landscape concept provides an explanation for the equality of thermal and dielectric relaxation times. The equality itself is once more demonstrated on the basis of literature data for salol.
\end{abstract}

\maketitle

\section{Introduction}

Broadband dielectric spectroscopy \cite{loidl} is the most versatile method to study the flow process in molecular liquids. However, the quantitative nature of the relation between dielectric signal (i.e. molecular reorientation) and shear flow is not yet clear. The classical Debye picture and its extension to viscoelasticity \cite{gemant,dimarzio} considers the molecule as a small sphere with a hydrodynamic radius $r_H$ immersed in the viscoelastic liquid. It predicts a slow dielectric decay, about a factor of fifty slower than the mechanical shear stress decay. The measured dielectric decay is on the average a factor of ten faster than this Debye prediction \cite{boettcher}. A thorough quantitative analysis of dielectric and shear data in seven glass formers \cite{niss,jakobsen} showed a general qualitative agreement with the Gemant-DiMarzio-Bishop extension \cite{gemant,dimarzio} of the Debye scheme, but a rather poor quantitative fit. Very recently, an alternative to the Debye model and its extensions was proposed \cite{bow}, which gave a much better fit for glycerol \cite{loidl,rossler} and propylene carbonate \cite{loidl2} data.

The proposal was based on the growing evidence for the identity of thermal and dielectric relaxation functions \cite{birge,ngai,donth,bow} (dielectric hole burning experiments \cite{jeffrey,weinstein} even suggest that dynamically distinct domains in the liquid are associated with a time constant characterizing both the dielectric and the thermal behavior).

The present paper begins with a reminder of the textbook introduction to dielectrics \cite{boettcher1} in terms of the decay of the dielectric polarization after a switch-off of the electric field. In the Debye scheme, the decay of the dielectric polarization is due to the rotational diffusion of the molecules with a Debye-Stokes-Einstein diffusion constant. But one can easily generalize the formalism to other decay mechanisms. In particular, we consider the recent proposal \cite{bow} of an initial and a retarded part of the flow process. We show that one must expect such a division in the energy landscape picture. The thermal relaxation sees both the initial and the retarded final part of the process.  The equality of thermal and dielectric relaxation times is understandable in terms of an energy landscape argument. The equality itself is checked for salol, using literature data.

\section{Theoretical basis}

In SI units, the electrostatic equations read
\begin{equation}
    D_0=\epsilon_0\epsilon(0)E_0=\epsilon_0E_0+P_0,
\end{equation}
where $\epsilon_0$ is the vacuum permittivity, $E_0$ is the electric field, $D_0$ is the displacement field and $P_0$ is the polarization per unit volume. $E_0$, $D_0$ and $P_0$ are vectors, so the static dielectric constant $\epsilon(0)$ is a second rank tensor. Here, however, we limit ourselves to isotropic liquids or glasses, so $\epsilon(0)$ reduces to a scalar. Similarly, we neglect the conductivity contribution, assuming that it can be simply subtracted.

The electrostatic equations are generalized to frequency-dependent equations \cite{boettcher1}
\begin{equation}
    D(\omega)=\epsilon_0\epsilon(\omega)E(\omega)=\epsilon_0E(\omega)+P(\omega)
\end{equation}
with frequency-dependent fields and dielectric constant, respectively. Thus one has
\begin{equation}\label{epsom}
    \epsilon(\omega)=1+\frac{1}{\epsilon_0}\frac{P(\omega)}{E(\omega)}.
\end{equation}

\begin{figure}[b]
\hspace{-0cm} \vspace{0cm} \epsfig{file=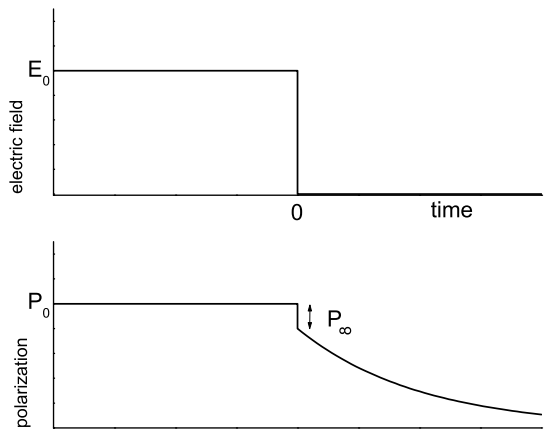,width=8
cm,angle=0} \vspace{0cm}\caption{Polarization decay after switch-off of the electric field at time zero.}
\end{figure}

The next step introduces a useful simplification for $\epsilon_\infty$, the dielectric constant at infinite frequency. Since one is mainly interested in the relaxation in the frequency range below 100 GHz, one considers all processes above this frequency as immediate processes. This includes the electronic polarizability (time scale 10$^{-15}$ seconds) as well as the vibrational contributions (time scale 10$^{-12}$ seconds), so
\begin{equation}
    \epsilon_\infty=n^2+\Delta\epsilon_{vib},
\end{equation}
where $n$ is the refractive index and $\Delta\epsilon_{vib}$ is the contribution of the vibrations (molecular librations) to the dielectric constant. Since this contribution is nonzero, one expects $\epsilon_\infty>n^2$ in a reasonable fit (this is in fact a main problem of the Gemant-DiMarzio-Bishop extension \cite{gemant,dimarzio} of the Debye scheme \cite{niss}).

Consider an electric field $E_0$ staying constant from $t=-\infty$ until $t=0$. At time zero, the field is switched off (Fig. 1). The Fourier transform of this field reads
\begin{equation}\label{eom}
    E(\omega)=-\frac{i}{\omega}E_0.
\end{equation}

For such a field, the polarization has a constant value
\begin{equation}\label{p0}
    P_0=(\epsilon(0)-1)\epsilon_0E_0
\end{equation}
for times $t<0$. At $t=0$, the polarization drops instantaneously to $P_0-P_\infty$, where
\begin{equation}\label{pinf}
    P_\infty=(\epsilon_\infty-1)\epsilon_0E_0
\end{equation}

At $t>0$, $P(t)$ decays with
\begin{equation}
    P(t)=(P_0-P_\infty)\Phi(t),
\end{equation}
where $\Phi(t)$ is a function which begins with 1 at time zero and drops to zero at infinite time.

If one compares this dielectric polarization decay with the mechanical shear stress decay, one finds a fundamental difference. After switching off the electric field, the polarization decays without any externally applied force. In the mechanical case, one applies a small external shear strain at time zero, keeps the strain constant and one observes the decay of the shear stress. The mechanical shear stress decay is described in linear response by a time-dependent shear modulus $G(t)$. The dielectric case is different. In order to describe it in terms of a time-dependent dielectric constant, one needs to go to the opposite case of Fig. 1, switching the electric field {\it on} at time zero and watching the rise of $\epsilon(t)$ from $\epsilon_\infty$ to $\epsilon_0$ with the function $1-\Phi(t)$. As a consequence, if one wants to compare decay with decay in order to see which quantity decays faster, one needs to compare the mechanical modulus with the dielectric susceptibility and not with any dielectric modulus, at variance with a recent proposal \cite{niss}. We come back to this point after the treatment of the Gemant-DiMarzio-Bishop relation \cite{gemant,dimarzio}.

The Fourier transform $P(\omega)$ of $P(t)$ is given by
\begin{equation}\label{pom}
    P(\omega)=-\frac{i}{\omega}P_0+(P_0-P_\infty)\emph{F}[\Phi(t>0)],
\end{equation}
where $\emph{F}[\Phi(t>0)]$ denotes the Fourier transform of the decay function $\Phi(t)$.

Inserting eqs. (\ref{eom}) and (\ref{pom}) into eq. (\ref{epsom}), one gets
\begin{equation}\label{epsom2}
    \epsilon(\omega)=1+\frac{1}{\epsilon_0}\left(\frac{P_0}{E_0}+\frac{P_0-P_\infty}{E_0}i\omega\emph{F}[\Phi(t>0)]\right).
\end{equation}

The relation between $\Phi(\omega)$ and $\Phi(t)$ is the same as the one between $G(\omega)$ and $G(t)$
\begin{equation}
\Phi(\omega)\equiv-i\omega F[\Phi(t>0)]=\omega\int_0^\infty\Phi(t)(\sin\omega t+i\cos\omega t)dt.
\end{equation}
$\Phi(\omega)$ is a complex function which is 1 for infinite frequency and decreases to zero as the frequency goes to zero.

With this definition, and inserting eqs. (\ref{p0}) and (\ref{pinf}) into eq. (\ref{epsom2}), we arrive at the final result
\begin{equation}\label{fin}
    \frac{\epsilon(\omega)-\epsilon_\infty}{\epsilon(0)-\epsilon_\infty}=1-\Phi(\omega).
\end{equation}

The Debye decay mechanism is the rotational diffusion of the molecules, with the diffusion constant given by the Debye-Stokes-Einstein equation
\begin{equation}
D_{trans}=\frac{k_BT}{6\pi\eta r_H}=\frac{4}{3}r_H^2D_{rot},
\end{equation}
where $D_{trans}$ is the translational diffusion constant of the molecule and $D_{rot}$ is its rotational diffusion constant. $r_H$ is the hydrodynamic radius of the molecule (note that the hydrodynamic Debye-Stokes-Einstein equation is in principle derived for larger objects than a single molecule). For continuous rotational diffusion, the relaxation time for the Legendre polynomials is
\begin{equation}
    \tau_{L,rot}=\frac{1}{L(L+1)D_{rot}},
\end{equation}
where $L$ is the order of the Legendre polynomial. For the dielectric signal,
$L=1$, one obtains the Debye relaxation time
\begin{equation}\label{taud}
    \tau_D=\frac{4\pi\eta r_H^3}{k_BT}.
\end{equation}
The decay function for the dielectric polarization is a single exponential (Debye process)
\begin{equation}
    \Phi(t)=\exp(-t/\tau_D)
\end{equation}
which leads to the well-known Debye equation
\begin{equation}\label{debye}
        \frac{\epsilon(\omega)-\epsilon_\infty}{\epsilon(0)-\epsilon_\infty}=\frac{1}{1+i\omega\tau_D}.
\end{equation}

To extend the Debye equation to viscoelasticity \cite{gemant,dimarzio}, one replaces the static viscosity $\eta$ by a frequency-dependent function $\eta(\omega)$, which is in turn related to the frequency-dependent shear modulus $G(\omega)$
\begin{equation}\label{etaom}
    i\omega\eta(\omega)=G(\omega)\equiv G_\infty g(\omega).
\end{equation}
Here $G_\infty$ is the infinite frequency shear modulus and $g(\omega)$ is a normalized function like $\Phi(\omega)$, going from zero to 1 with increasing frequency.

The ratio between viscosity and infinite frequency shear modulus defines the Maxwell time
\begin{equation}\label{taumax}
    \tau_M=\frac{\eta}{G_\infty},
\end{equation}
the characteristic shear relaxation time.

If we replace the viscosity $\eta$ in the $\tau_D$ of the right hand side of the Debye equation by the expression of eq. (\ref{etaom}), we get the Gemant-DiMarzio-Bishop expression
\begin{equation}\label{gdb1}
    \frac{1}{1+(4\pi G_\infty r_H^3/k_BT)g(\omega)}\equiv\frac{1}{1+c_rg(\omega)},
\end{equation}
which contains the dimensionless ratio $c_r$ between the Debye relaxation time $\tau_D$ and the Maxwell time $\tau_M$.

The Gemant-DiMarzio-Bishop expression, eq. (\ref{gdb1}), does not approach zero for infinite frequency as the Debye expression in eq. (\ref{debye}), because the molecule is still able to turn in a balance between the torques exerted by the  electric field and the finite restoring force of the sheared elastic medium, respectively. In fact, one has in this case
\begin{equation}
\Delta\epsilon_{vib}=\frac{\epsilon(0)-n^2}{1+c_r}.
\end{equation}
Thus $\Delta\epsilon_{vib}$ is already taken into account, and the Gemant-DiMarzio-Bishop equation \cite{gemant,dimarzio} takes the form
\begin{equation}\label{gdb}
        \frac{\epsilon(\omega)-n^2}{\epsilon(0)-n^2}=\frac{1}{1+c_rg(\omega)}.
\end{equation}

With a hydrodynamic radius of 0.2 nm, a $G_\infty$ of 2 GPa and a temperature of 200 K one calculates $c_r\ =\ 50$, so one expects a small vibrational component and nearly two decades difference between $\tau_D$ and the Maxwell time. The first of these expectations is in fact found, but the second fails by an average factor of ten \cite{boettcher,chang,niss,bow}.

Note that the failure is only in the factor and not in the temperature dependence. Unlike the translational diffusion, which decouples from the viscosity below some critical temperature \cite{chang,chang2,ediger}, the rotational relaxation time follows essentially the temperature dependence of the viscosity over the whole temperature range.

On the basis of the Gemant-DiMarzio-Bishop equation (\ref{gdb}), it has been recently argued \cite{niss} that one should compare $G(\omega)$ with $1/(\epsilon(\omega)-n^2)$, the so-called "rotational modulus". This is true if the equation is valid; it is not true if one has a different decay mechanism of the dielectric polarization. Take, for instance, the Debye model of eq. (\ref{debye}). The true relaxation peak lies at $1/\tau_D$ in $\epsilon(\omega)$ as it should, but shifts with varying $\Delta\epsilon_{vib}$ in the rotational modulus $1/(\epsilon(\omega)-n^2)$.

\begin{figure}[b]
\hspace{-0cm} \vspace{0cm} \epsfig{file=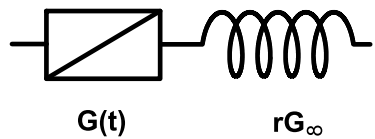,width=6cm,angle=0} \vspace{0cm}\caption{Spring model for the decay of the structural potential energy of an undercooled liquid.}
\end{figure}

An alternative to the Debye-Gemant-DiMarzio-Bishop scheme is a recent proposal \cite{bow} for the structural relaxation in the energy landscape \cite{goldstein} (for a review of the energy landscape concept see \cite{diezemann}) The proposal is based on the experimental finding \cite{donth} of two time scales in the undercooled liquid. The faster one is the shear stress decay, described by $g(\omega)$. The slower one is the decay of the structural potential energy, seen in dynamic heat capacity measurements or in transient grating experiments \cite{yn}.

To understand this behavior, let us assume that the shear stress decay occurs via thermally activated processes in the energy landscape \cite{goldstein,diezemann}, without specifying the exact nature of the thermally activated processes. They might be "flow by shoving" \cite{dyre} or some other mechanism \cite{buth}; for our purpose, it suffices to assume that the shear decay happens by repeated passages of the system from one minimum of the energy landscape to another. The energy landscape picture provides a natural explanation for the dynamical heterogeneity seen in numerous experiments \cite{richert}. One can show that an initial and a retarded part of the flow process follow from the energy landscape concept under reasonable assumptions.

To see this, consider two structural rearrangements occurring in different parts of the sample. Each of them can be characterized by an asymmetric double-well potential in the corresponding configurational coordinate, leading from one potential energy minimum to the other. Since the two structural rearrangements are far apart on a microscopic scale, the two minima of the first structural rearrangement change only slightly by a jump in the other double-well.

In a macroscopic sample, there is a very large number of possible structural rearrangements. Each of them is a local process, and each of them influences every other one. 

The consideration shows that one has two relaxation times in each double-well potential, a lifetime of the population of the two minima and a lifetime of the double-well potential itself. For the latter one, the Maxwell time is crucial, because it sets a lower boundary for the time scale on which the core region of the relaxation can change its shape. If the core region can change its shape, the core relaxation can disappear or make way for a different relaxation. This relaxation life time is not determined by a single jump, but by the joint effect of all the other relaxations in the sample, a diffusional motion through phase space. The dielectric data seem to see this final relaxation time. In fact, the dielectric relaxation time in glycerol and OTP is close to the spin-echo relaxation time at the first sharp diffraction peak \cite{bow}, the decay time of the short range order.

Naturally, the energy landscape itself does not change in time; nevertheless, we believe that the concept of a decay of a local double-well potential, for simplicity assumed to be the same for each double-well in the system, catches an essential feature of the motion of the system in the fixed energy landscape. Qualitatively, such a picture is also compatible with NMR findings \cite{diezemann2} in toluene and glycerol, showing a small number of larger-angle rotational jumps (attributable to the hopping part) and a larger number of small-angle rotational jumps of the molecules (attributable to the diffusional part).

The final part of the equilibration is not (or maybe only partially) seen in the stress decay, because it can only happen
after the stress decay. As long as there is still some appreciable shear rigidity, higher and higher barriers are jumped over in order to equilibrate the remaining average stress to zero. This mechanism determines the Maxwell time, and a corresponding Maxwell barrier. Roughly speaking, the part of the motion through phase space due to the initial change of conditions changes from hopping to diffusion after the Maxwell time. If an energy barrier is higher than the Maxwell barrier, it is not jumped over, but it flows away.

Here, we follow ref. \cite{bow} in assuming that the retarded part of the process can be described as the decay of an energy stored in a harmonic spring $rG_\infty$ in series with a time-dependent spring $G(t)$ (see Fig. 2). The lower $r$, the more retarded is this final stage of the equilibration. $r$ is the ratio of the energy stored in $G(t)$ to the one stored in the harmonic spring, which does only decay by flowing over into the spring $G(t)$.

Note that not the structural potential energy itself decays, but its difference to an average temperature-dependent value
determined by the entropy of the inherent structure of the energy landscape. Our assumptions imply that the difference between the potential structural energy of a given minimum of the energy landscape and the average value has always the ratio $r$ between its harmonic part and its long-range stress component.

In the decay of the structural potential energy of the undercooled liquid, one expects to see both the initial hopping and the final diffusion. The previous paper \cite{bow} took only this second retarded action into account. But naturally one has to take the initial process into account as well. Each jump in an asymmetric double-well changes the structural potential energy. This forces one to introduce an additional dimensionless parameter $f$, the fraction of the structural potential energy which equilibrates via the initial hopping. In a globally connected energy landscape \cite{diezemann}, this parameter $f$ would be one, but in a real energy landscape it is expected to be considerably smaller.

We assume that this initial part has the same time dependence as the shear modulus $G(t)$. Then the Fourier transform of the normalized decay function of the structural potential energy is
\begin{equation}
    \Phi(\omega)=fg(\omega)+(1-f)\frac{1+r}{1+r/g(\omega)},
\end{equation}
where $g(\omega)=G(\omega)/G_\infty$.

If the dielectric polarization has indeed the same time dependence as the structural potential energy difference, we have to insert this $\Phi(\omega)$ into eq. (\ref{fin}) and obtain
\begin{equation}\label{eps}
\frac{\epsilon(\omega)-\epsilon_\infty}{\epsilon(0)-\epsilon_\infty}=f(1-g(\omega))+(1-f)\frac{1-g(\omega)}{1+g(\omega)/r}.
\end{equation}

Experimentally, one finds the loss peak of the dielectric constant close to the one of the heat capacity \cite{bow,birge,ngai,dixon,donth}. In fact, one can argue that the electric field introduces an imbalance between the different structural realizations of the undercooled liquid at the given temperature. If one mirrors the sample at a plane perpendicular to the electric field, one gets a state with the same structural potential energy, but with opposite electric dipole moment. These two states have the same energy in the absence of the electric field, but a different energy in the electric field. Thus the switch-off of the field leaves an imbalance in the structural entropy. It should equilibrate by the same mechanism as the imbalance introduced by a temperature jump. On the basis of this argument, one understands the coincidence of the dielectric relaxation time with the dynamic heat capacity and the transient grating one.

If the decay mechanism of the dielectric polarization is not the diffusion of a single molecule, but the passage from one of the minima of the energy landscape to another, one needs no longer distinguish between the internal field at the molecule and the externally applied field, because the externally applied field acts directly on the electric dipole moment difference of the two minima; all interaction effects are already included in this dipole moment difference. The complicated many-body problem of a system of interacting electric dipoles \cite{boettcher1} enters only via the temperature dependence of $\epsilon(0)$.

It is interesting to consider the limiting cases of very small and very large $r$ in the retarded term of eq. (\ref{eps}). For very small $r$, the denominator dominates the behavior and we approximate the Gemant-DiMarzio-Bishop case of eq. (\ref{gdb}). For very large $r$, the harmonic spring is like a rigid connection and the relaxation peaks of $G(\omega)$ and $\epsilon(\omega)$ become identical. Note that this limit can never be attained by the Gemant-DiMarzio-Bishop extension of the Debye scheme. There, it corresponds to the limit of very small $c_r$, in which the relaxation peak disappears, leaving only the vibrational component.

\section{Dielectric, thermal and rotational relaxation times in salol}

For the comparison of dielectric and thermal relaxation times, we add a new example, salol (glass temperature 213 to 217 K, mode coupling critical temperature 255 to 265 K), to the three examples glycerol, propylene carbonate and OTP discussed in the previous work \cite{bow}. All measured relaxation times are compared to the Maxwell time, calculated from fits of the infinite frequency shear modulus and of the viscosity. The infinite frequency shear modulus is taken from light scattering Brillouin data. As it turns out, it is better to use longitudinal sound wave data, relating the transverse sound velocity $v_t$ to the longitudinal $v_l$ by $v_l/v_t\approx 1.8$. The mistake of this approximation is small on the scale of factors of ten considered here. The advantages are (i) the longitudinal infinite frequency sound velocity is much more easily determined (ii) one has data up to the highest temperatures. To take an example, $G_\infty$ from transverse Brillouin OTP data \cite{dreyf} extrapolates to zero at 348 K (not at 308 K as stated erroneously in the previous paper), while one still has data to compare up to 380 K.

In salol, the density follows the relation \cite{enright}
\begin{equation}
    \rho=1451.6-0.857T
\end{equation}
 with $\rho$ in $kg/m^3$ and $T$ in Kelvin. The longitudinal infinite frequency sound velocity \cite{dreyf2}
\begin{equation}
    v_l=2400 \left( \frac{T_g}{T}\right)^{0.88}\ m/s
\end{equation}
with the glass transition temperature $T_g=218 K$. With our recipe for the ratio $v_l/v_t$, this yields
\begin{equation}
    G_\infty=2.2\left(\frac{T_g}{T}\right)^{1.96}\ GPa.
\end{equation}

The viscosity \cite{jantsch,laughlin,cukierman} is parametrized in terms of two overlapping Vogel-Fulcher laws
\begin{equation}\label{vfth}
\log\eta=\log\eta_{0i}+\frac{B_i}{T-T_{0i}}
\end{equation}
with $i=1$ and $i=2$, respectively.
The first of these two is valid below a temperature $T_1$, the second above a temperature $T_2\leq T_1$. Between $T_2$ and $T_1$, one takes a linear interpolation between the two to ensure continuity. For salol, $\log\eta_{01}=-36.9$ and $\log\eta_{02}=-3.68$ ($\eta$ in Pas), $B_1=7500$ K and $B_2=151$ K, $T_{01}=59$ K and $T_{02}=225$ K, $T_1=256$ K and $T_2=248$ K.

\begin{figure}[b]
\hspace{-0cm} \vspace{0cm} \epsfig{file=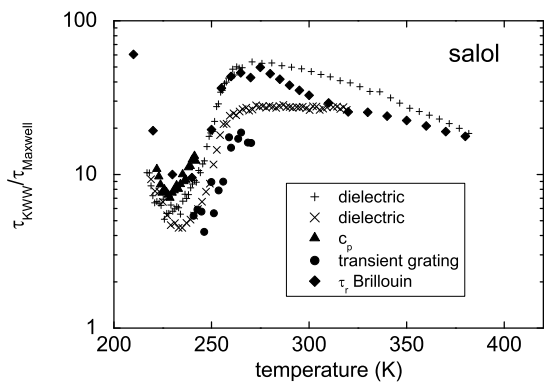,width=8cm,angle=0} \vspace{0cm}\caption{Kohlrausch-Williams-Watts relaxation times in salol, normalized to the Maxwell time as described in the text. Symbols: pluses dielectric \cite{dixon}; crosses dielectric \cite{stickel}; full triangles heat capacity \cite{dixon}; full diamonds rotational relaxation times extracted from Brillouin light scattering \cite{zhang}; full circles transient grating data \cite{yn2}.}
\end{figure}

Figure 3 compares dielectric \cite{dixon,stickel}, heat capacity \cite{dixon} and transient grating \cite{yn2} relaxation times to rotational relaxation times extracted from longitudinal Brillouin light scattering data \cite{zhang}. The times were either recalculated \cite{patt} or refitted in terms of a Kohlrausch function $exp(-(t/\tau_{KWW})^\beta)$.

Fig. 3 contains no mechanical shear relaxation times, but one knows that these have to lie below the Maxwell time, because for a shear modulus following the Kohlrausch function $exp(-(t/\tau_{shear})^\beta)$
\begin{equation}\label{taukww}
\frac{\tau_{shear}}{\tau_{Maxwell}}=\frac{\beta}{\Gamma(1/\beta)}.
\end{equation}
Usually, $\beta$ lies between 0.4 and 0.6, so the ratio should
be between one third and two thirds; the shear Kohlrausch relaxation time should be a factor 1.5 to 3 shorter
than the Maxwell time.

Fig. 3 shows once again that the rotational times keep close to the entropy relaxation times in the heat capacity and in the transient grating measurements, thus supporting the proposed energy landscape mechanism for the relaxation of the molecular orientation. In principle, the transient grating measurements should also be evaluated in the Ansatz of Pick and Dreyfus \cite{dreyf3,dreyf4}, because the signal comes from a grating of both temperature and local molecular orientation. However, an experimental separation \cite{yn3} of the contributions demonstrated again the equality of the two relaxation functions within experimental error.

Note that the method developed by Dreyfus and Pick \cite{dreyf3,dreyf4} to determine rotational relaxation times from light scattering Brillouin data seems to work remarkably well. According to their Ansatz, one should see the second Legendre polynomial.  For continuous rotational diffusion, their relaxation times should be a factor of three smaller than the dielectric ones. One does not see this factor of three; the two sets of data are remarkably close to each other over ten decades in relaxation time, even making the same wiggles compared to the Maxwell time. In the energy landscape mechanism, one does not expect the factor three; there, it depends on the jump angle distribution which can be different for different glass formers \cite{diezemann,diezemann2}.

The marked temperature dependence of the ratio between rotational relaxation time and Maxwell time indicates a temperature dependence of the retardation parameter $r$ in eq. (\ref{eps}). The rise at low temperature is also seen in glycerol and in propylene carbonate \cite{bow}. The salol kinks at 260 and 240 K have been also seen in the "Stickel plot" \cite{stickel} of the dielectric data (a plot of the inverse square root of the temperature derivative of the peak frequency versus temperature, where one sees the deviations from a perfect Vogel-Fulcher behavior).

From the point of view of the mode coupling theory \cite{mct}, the kink at 260 K should mark the crossover from an energy landscape behavior below $T_c$ to simple liquid behavior above. But Fig. 3 shows that the separation of time scales persists into the liquid domain in salol. This time scale splitting is not the two-stage scenario of the mode-coupling theory, because both time scales move together with the Maxwell time. In fact, in ref. \cite{brodin} this time scale splitting was also observed for propylene carbonate and discussed in terms of the mode coupling theory. The
$\alpha$-process of the theory was not attributed to the slower, but to the faster process.

\section{Summary}

The textbook relation between dielectric polarization decay and dielectric constant is applied to the simple Debye case, to its extension to viscoelasticity and to a newly proposed energy landscape mechanism. 

Data in salol show once again the equality of dielectric and structural relaxation times, supporting the energy landscape mechanism.

We thank Catherine Dreyfus and Robert Pick for helpful discussions.


\begin{thebibliography}{99}
\bibitem{loidl} U. Schneider, P. Lunkenheimer, R. Brand and A.
Loidl, J. Non-Cryst. Solids {\bf 235-237}, 173 (1998)
\bibitem{gemant} A. Gemant, Trans. Faraday Society {\bf 31}, 1582
(1935)
\bibitem{dimarzio} E. A. DiMarzio and M. Bishop, J. Chem. Phys.
{\bf 60}, 3802 (1974)
\bibitem{boettcher} C. J. F. B\"ottcher and P. Bordewijk, {\it Theory of Electric
Polarization}, Volume II, Elsevier, Amsterdam 1978, Table 32, p. 212
\bibitem{niss} K. Niss, B. Jakobsen and N. B. Olsen, J. Chem. Phys.
{\bf 123}, 234510 (2005)
\bibitem{jakobsen}B. Jakobsen, K. Niss
and N. B. Olsen, J. Chem. Phys. {\bf 123}, 234511 (2005)
\bibitem{bow} U. Buchenau, M. Ohl and A. Wischnewski, J. Chem. Phys.
{\bf 124}, 094505 (2006)
\bibitem{rossler} S. Adichtchev, T. Blochowicz, C. Tschirwitz, V.
N. Novikov and E. A. R\"ossler, Phys. Rev. E {\bf 68}, 011504
(2003)
\bibitem{loidl2} U. Schneider, P. Lunkenheimer, R. Brand and A.
Loidl, Phys. Rev. E {\bf 59}, 6924 (1999)
\bibitem{boettcher1} C. J. F. B\"ottcher and P. Bordewijk, {\it Theory of Electric
Polarization}, Volume II, Elsevier, Amsterdam 1978, p. 5 ff.
\bibitem{birge} N. O. Birge and S. R. Nagel, Phys. Rev. Lett.
{\bf 54}, 2674 (1985); N. O. Birge, Phys. Rev. B {\bf 34}, 1631
(1986)
\bibitem{ngai} K. L. Ngai and R. W. Rendell, Phys. Rev. B {\bf 41}, 754 (1990)
\bibitem{donth} K. Schr\"oter and E. Donth, J. Non-Cryst. Solids
{\bf 307-310}, 270 (2002)
\bibitem{jeffrey}K. R. Jeffrey, R. Richert and K. Duvvuri, J. Chem. Phys. {\bf 119}, 6150 (2003)
\bibitem{weinstein} S. Weinstein and R. Richert, J. Chem. Phys. {\bf 123}, 224506 (2005)
\bibitem{chang} I. Chang and H. Sillescu, J. Chem. Phys. {\bf 101}, 8794
(1997) and further references therein
\bibitem{chang2} I. Chang, F. Fujara, B. Geil, G. Heuberger, T. Mangel and
H. Sillescu, J. Non-Crystalline Solids {\bf 172-174}, 248 (1994)
\bibitem{ediger} M. Cicerone, F. R. Blackburn and M. D. Ediger, J. Chem. Phys. {\bf 102},
471 (1995)
\bibitem{yn} Y. Yang and K. A. Nelson, J. Chem. Phys.
{\bf 103}, 7722 (1995)
\bibitem{goldstein}M. Goldstein, J. Chem. Phys. {\bf 51}, 3728
(1969)
\bibitem{diezemann} G. Diezemann, J. Chem. Phys. {\bf 107}, 10112
(1997)
\bibitem{dyre} J. C. Dyre, N. B. Olsen and T. Christensen, Phys. Rev. B {\bf 53}, 2171 (1996)
\bibitem{buth} U. Buchenau, J. Phys.: Condens. Matter {\bf 15}, S955 (2003)
\bibitem{richert} R. Richert, J. Phys.: Condens. Matter {\bf 14}, R703 (2002)
\bibitem{diezemann2} G. Diezemann, R. B\"ohmer, G. Hinze and H. Sillescu, J. Non-Cryst. Solids {\bf 235-237}, 121 (1998)
\bibitem{dreyf} C. Dreyfus, A. Aouadi, J. Gapinski, M. Matos-Lopes,
W. Steffen, A. Patkowski and R. M. Pick, Phys. Rev. E {\bf 68}, 011204 (2003)
\bibitem{enright} G. D. Enright and B. P. Stoicheff, J. Chem. Phys. {\bf 64}, 3658 (1976)
\bibitem{dreyf2} C. Dreyfus, M. J. Lebon, H. Z. Cummins, J. Toulouse, B. Bonello and R. M. Pick, Phys. Rev. Lett. {\bf 69}, 3666 (1992)
\bibitem{jantsch} V. O. Jantsch, Z. Kristallogr. {\bf 108}, 185 (1956)
\bibitem{laughlin}W. T. Laughlin and D. R. Uhlmann, J. Phys. Chem. {\bf 76},
2317 (1972)
\bibitem{cukierman}M. Cukierman, J. W. Lane and D. R. Uhlmann, J. Chem. Phys. {\bf 59},
3639 (1973)
\bibitem{dixon} P. K. Dixon, Phys. Rev. B {\bf 42}, 8179 (1990)
\bibitem{stickel} F. Stickel, E. W. Fischer and R. Richert, J. Chem. Phys. {\bf 102}, 6251 (1995)
\bibitem{zhang} H. P. Zhang, A. Brodin, H. C. Barshilia, G. Q. Shen and R. Pick, Phys. Rev. E {\bf 70}, 011502 (2004)
\bibitem{yn2} Y. Yang and K. A. Nelson, Phys. Rev. Lett. {\bf 74}, 4883 (1995)
\bibitem{patt} C. P. Lindsey and G. D. Patterson, J. Chem. Phys.
{\bf 73}, 3348 (1980)
\bibitem{dreyf3} C. Dreyfus, A. Aouadi, R. M. Pick, T. Berger, A. Patkowski and W. Steffen, Europhys. Lett. {\bf 42}, 55 (1998)
\bibitem{dreyf4} R. M. Pick, T. Franosch, A. Latz and C. Dreyfus, Eur. Phys. J. B {\bf 31}, 217 (2003)
\bibitem{yn3} C. Glorieux, K. A. Nelson, G. Hinze and M. D. Fayer, J. Chem. Phys.
{\bf 116}, 3384 (2002)
\bibitem{mct} W. G\"otze and L. Sj\"ogren, Rep. Prog. Phys. {\bf 55},
241 (1992)
\bibitem{brodin} A. Brodin, M. Frank, S. Wiebel, G. Shen, J. Wuttke
and H. Z. Cummins, Phys. Rev. E {\bf 65}, 051503 (2002)
\end{thebibliography}
\end{document}